\documentclass[10pt]{iopart}
\usepackage{iopams}  
\usepackage{graphicx}
\usepackage{bm}
\usepackage{mathrsfs,amssymb}
\usepackage{cite}
\usepackage[colorlinks=true,linkcolor=blue,citecolor=blue,urlcolor=blue]{hyperref}

\newcommand{\ud}{\,\mathrm{d}}
\newcommand{\ket}[1]{|#1\rangle}
\newcommand{\bra}[1]{\langle#1|}
\newcommand{\ketbra}[2]{|#1\rangle\langle#2|}

\begin{document}

\title{Scalable cavity quantum electrodynamics system for quantum computing }


\author{Mohammad Hasan Aram and Sina Khorasani}

\address{Department of electrical engineering, Sharif university of technology, Tehran, Iran}
\ead{khorasani@sina.sharif.edu}
\vspace{10pt}
\begin{indented}
\item[]July 2015
\end{indented}

\begin{abstract}
We introduce a new scalable cavity quantum electrodynamics platform which can be used for quantum computing. This system is composed of coupled photonic crystal (PC) cavities which their modes lie on a Dirac cone in the whole super crystal band structure. Quantum information is stored in quantum dots that are positioned inside the cavities. We show if there is just one quantum dot in the system, energy as photon is exchanged between the quantum dot and the Dirac modes sinusoidally. Meanwhile the quantum dot becomes entangled with Dirac modes. If we insert more quantum dots into the system, they also become entangled with each other. 
\end{abstract}


%
\vspace{2pc}
\ioptwocol
%

\section{Introduction}

After about seventy years from Purcell's famous paper \cite{Purcell} which established cavity quantum electrodynamics (CQED), this field is still active and interesting for many researchers \cite{Walther}. This is primarily due to a concept called coupling constant. It is a criterion to measure the strength of atom-cavity interaction. Atom interacts with cavity through exchange of energy quanta or photon. The stronger this interaction is, the faster the exchange of photon occurs. So we can say coupling constant measures the rate of photon exchange between atom and cavity.

But why has this simple concept made this field so attractive? The answer is behind its role in quantum information and computation theory. We know quantum computers are powerful in solving some kind of problems. This is due to an inherent parallel processing power in them that originates from quantum physics. In order to use this capability of quantum physics we have to create entangled states between qubits of the quantum computer. In other words, if we save information on qubits which are not entangled, our computer has no advantage over its classical counterparts.
By increasing the rate of photon exchange between atom and cavity, we can create entangled state of them. To measure this rate we need a criterion, that is, we have to compare it with an amount to determine if it is high or not. There are three criteria as follows:

\begin{enumerate}
\item
Presence duration of atom inside the cavity. In some cavity quantum electrodynamics systems atoms stay inside the cavity for a short period of time. In these systems atoms enter the cavity from an aperture and after interaction exit from the other side. If energy exchange occurs in a period of time longer than atom presence duration ($T_e$), then the coupling constant is small and we say the coupling is weak.   

\item
Photon decay rate. We know there is no lossless cavity. This loss has many causes among them we can mention leakage through cavity walls and cavity walls absorption. If we show photon annihilation rate by $\xi$, photon exchange rate must be larger than $\xi$ to have strong coupling.

\item
Atom spontaneous emission rate. Another factor to be considered for measuring coupling strength is the time interval that atom can maintain photon before radiating it to vacuum modes outside the cavity via spontaneous emission. Spontaneous emission rate is shown by $\gamma$. So to have strong coupling we need photon exchange rate to be much larger than $\gamma$.
\end{enumerate}

The photon exchange rate between atom and cavity (coupling constant) is measured by Rabi frequency (g). So the last paragraph can be summarized as: If $g \gg (\xi , \gamma , 1/T_e )$ , then the atom-cavity system is in the strong coupling regime. Otherwise the coupling is weak. In many systems like the one analyzed here, quantum dots are used instead of real atoms or ions. Since quantum dots are always inside the cavity, we can assume $T_e \simeq \infty$. Hence in these systems we just need to compare $g$ with $\gamma$ and $\xi$ to determine the coupling strength. Usually photon decay rate is greater than atom spontaneous emission rate \cite{Mabuchi}, so it is usually sufficient to compare $g$ with $\xi$.

In strong coupling regime different phenomena occur. Among them is vacuum Rabi splitting in which upper atomic energy level splits into two close levels which results in two peaks in spontaneous emission spectrum of the atom that had only one otherwise \cite{Raizen4}.

Many reports of strong coupling attainment have been presented till now \cite{Ohta,Thon,Reithmaier,Lev4,Buck4,Aoki4,Hennessy4}. But in recent years some groups are trying to go beyond strong coupling and reach ultra-strong coupling regime \cite{Casanova4,Liberato4,Gunter4}. In this regime Rabi frequency is comparable with photon frequency. It is predicted atom behaves chaotically in this regime \cite{Alidoosty}.

CQED has many applications, but its capability to be used as a platform for quantum computing is much more attractive. Up to now different systems have been proposed to be used for quantum computing \cite{Ladd,Vandersypen,Steffen,Schoelkopf,Houck,DiCarlo}. All of these systems have some limitations and shortcomings. One of their serous limitations is their hard scalability. In most of these systems we can have only a few numbers of qubits. Here we propose a new system which does not have this limitation and can be used as an alternative for the old CQED systems in quantum computation. This system is composed of coupled photonic crystal resonators array (CPCRA) where each resonator is a cavity that can contain a quantum dot. The cavities are arranged such that a Dirac cone appears in the super crystal band structure.

In the remaining we first show how we can obtain Dirac cone by choosing appropriate position for the cavities, next we obtain Hamiltonian of the system that makes study of atom evolution due to interaction with Dirac cone modes possible.

\section{Dirac cone modes}

\begin{figure}[ht]
\centering
\includegraphics[width=\linewidth]{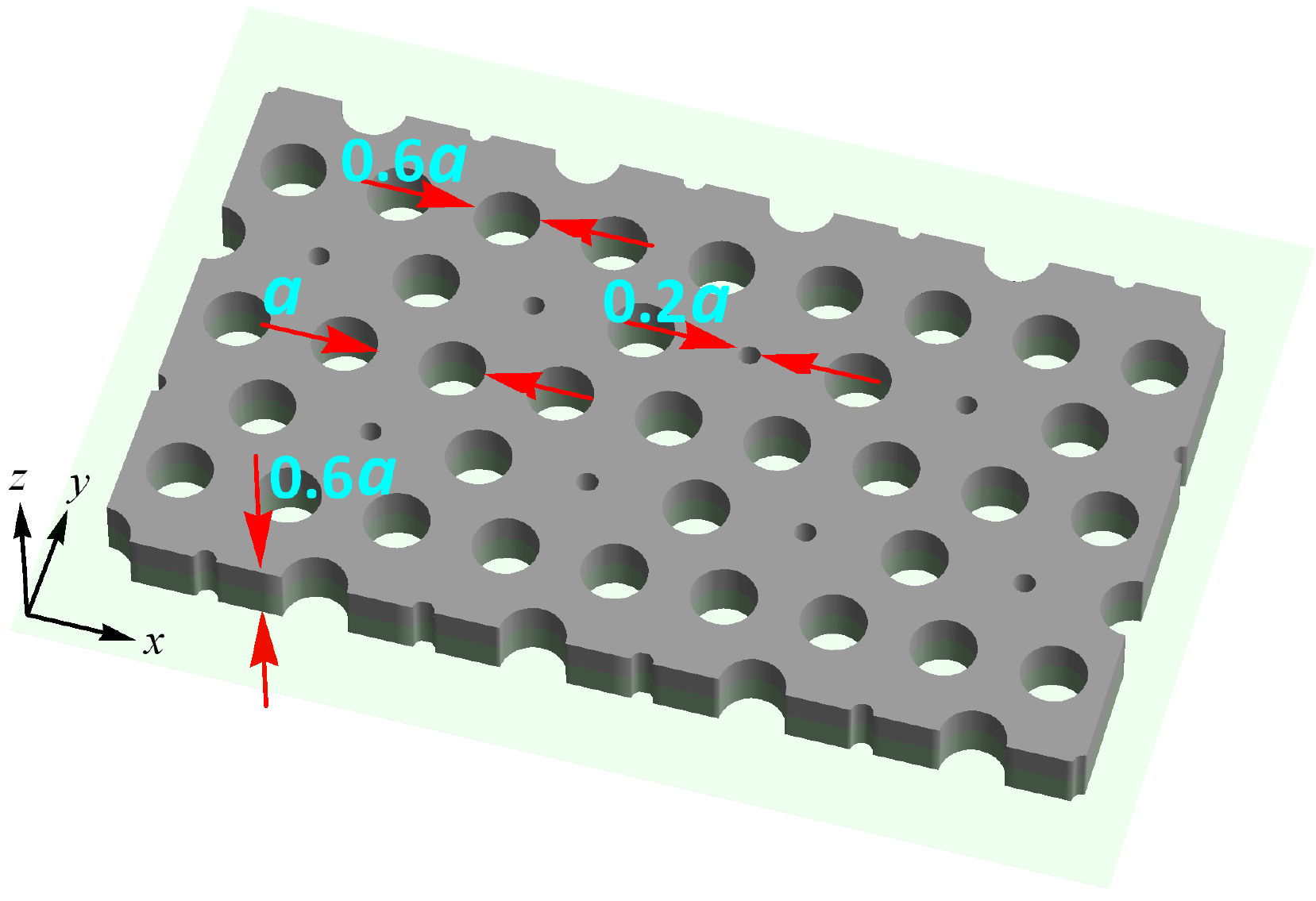}
\caption{\label{f1} Photonic crystal slab with Dirac point in band structure. Crystal dielectric is Silicon with $\epsilon_r=11.9$}
\end{figure}

In one of our previous works we showed how we can design a photonic crystal that has Dirac point in its transverse electric (TE) band structure \cite{Aram}. In this section we review it briefly. At first we designed a two dimensional (2D) crystal with Dirac point. We adopted the crystal lattice from graphene which has a honey comb lattice and Dirac point has already been observed in its band structure. So we chose a triangular lattice of air holes in a dielectric for the basis crystal and replaced Carbon atoms by cavities. Different patterns for cavities positions were tested to finally reach the pattern shown in figure \ref{f1}. We assumed the crystal to be made of Silicon with relative permittivity of $\epsilon_r =11.9$. 
Next we generalized our design to a photonic crystal slab. We used different methods to show Dirac point in its band structure but here we just explain tight-binding because we need it in subsequent sections. 

\begin{figure*}[ht!]
\centering
\includegraphics[width=0.25\linewidth]{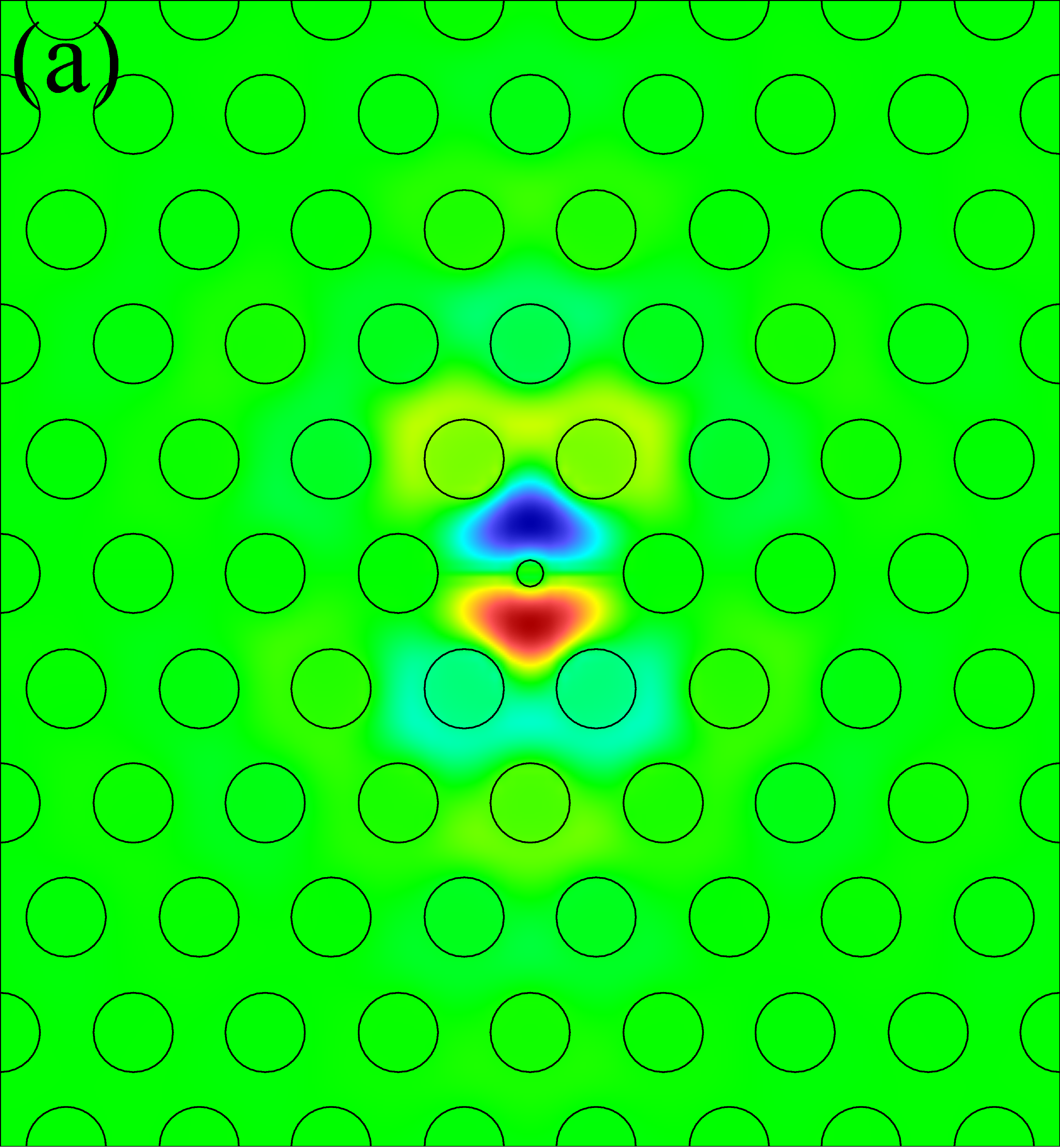} 
\includegraphics[width=0.25\linewidth]{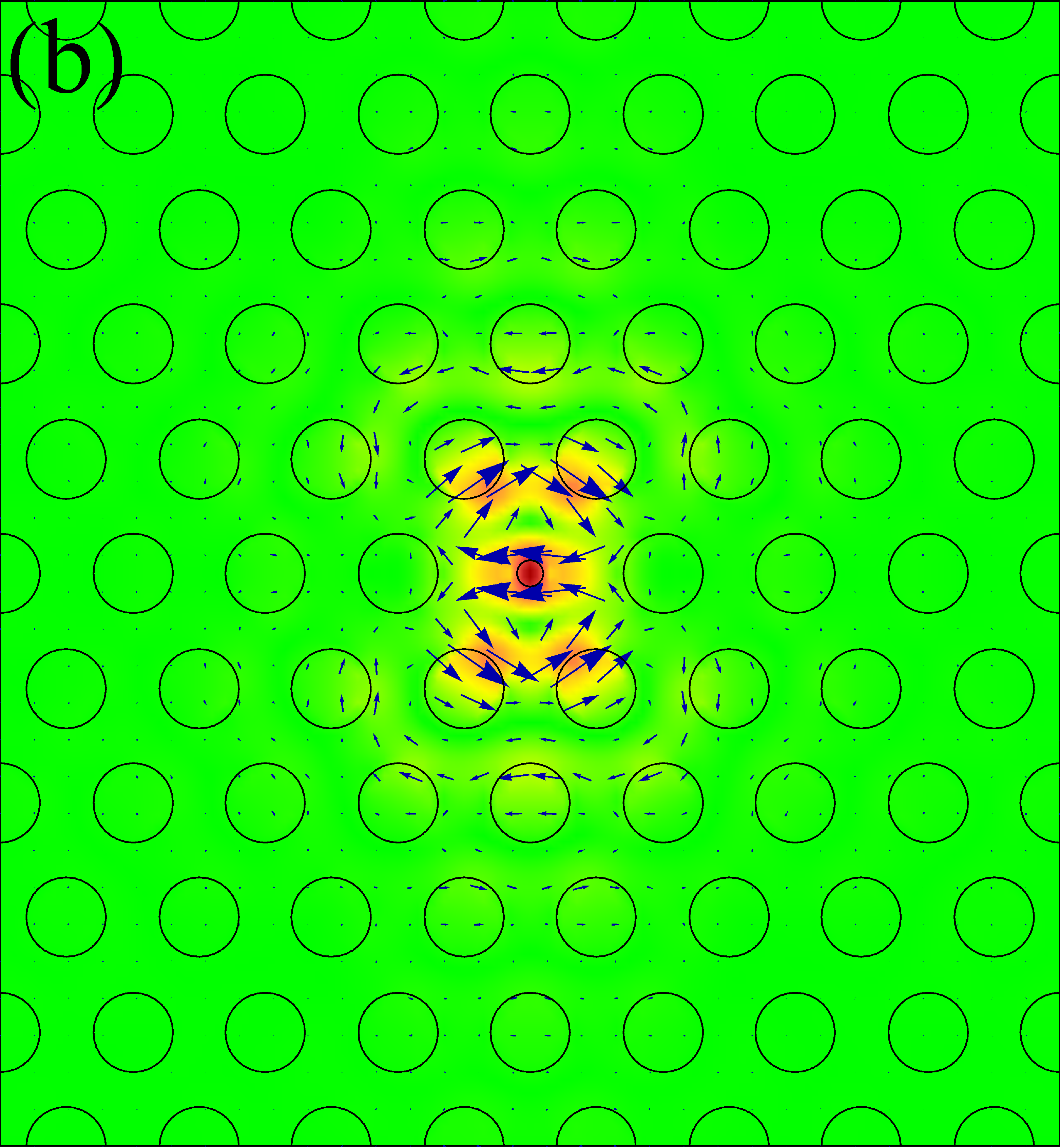} 
\includegraphics[width=0.25\linewidth]{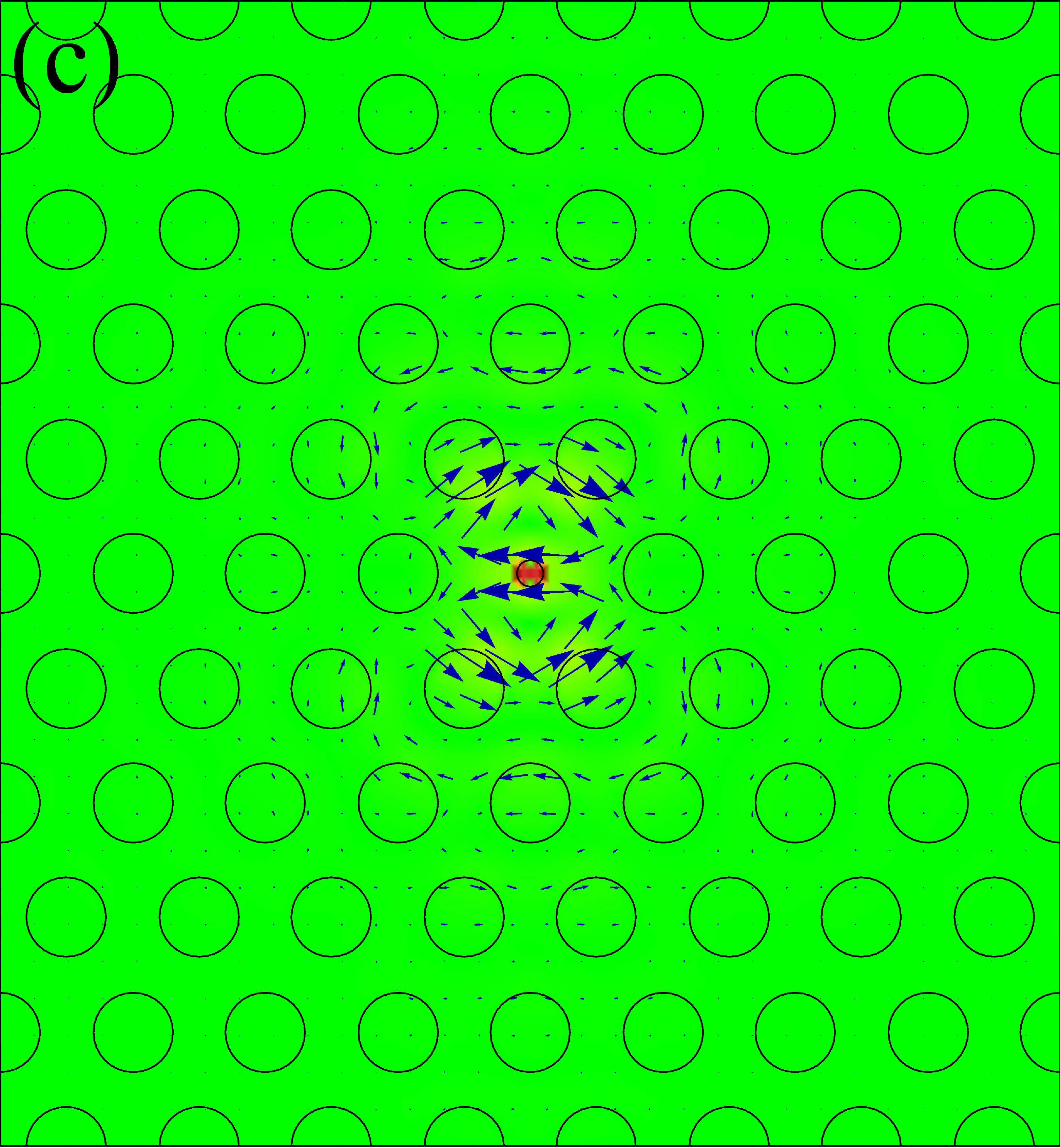} \hspace{1mm}
\includegraphics[width=0.03\linewidth]{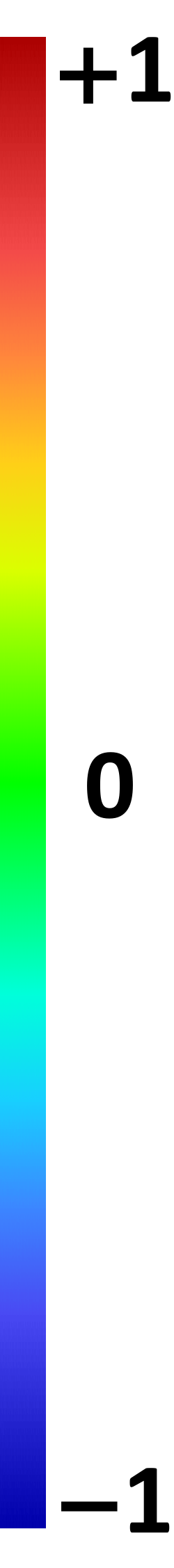} \\ \vspace{0.5 mm}
\includegraphics[width=0.25\linewidth]{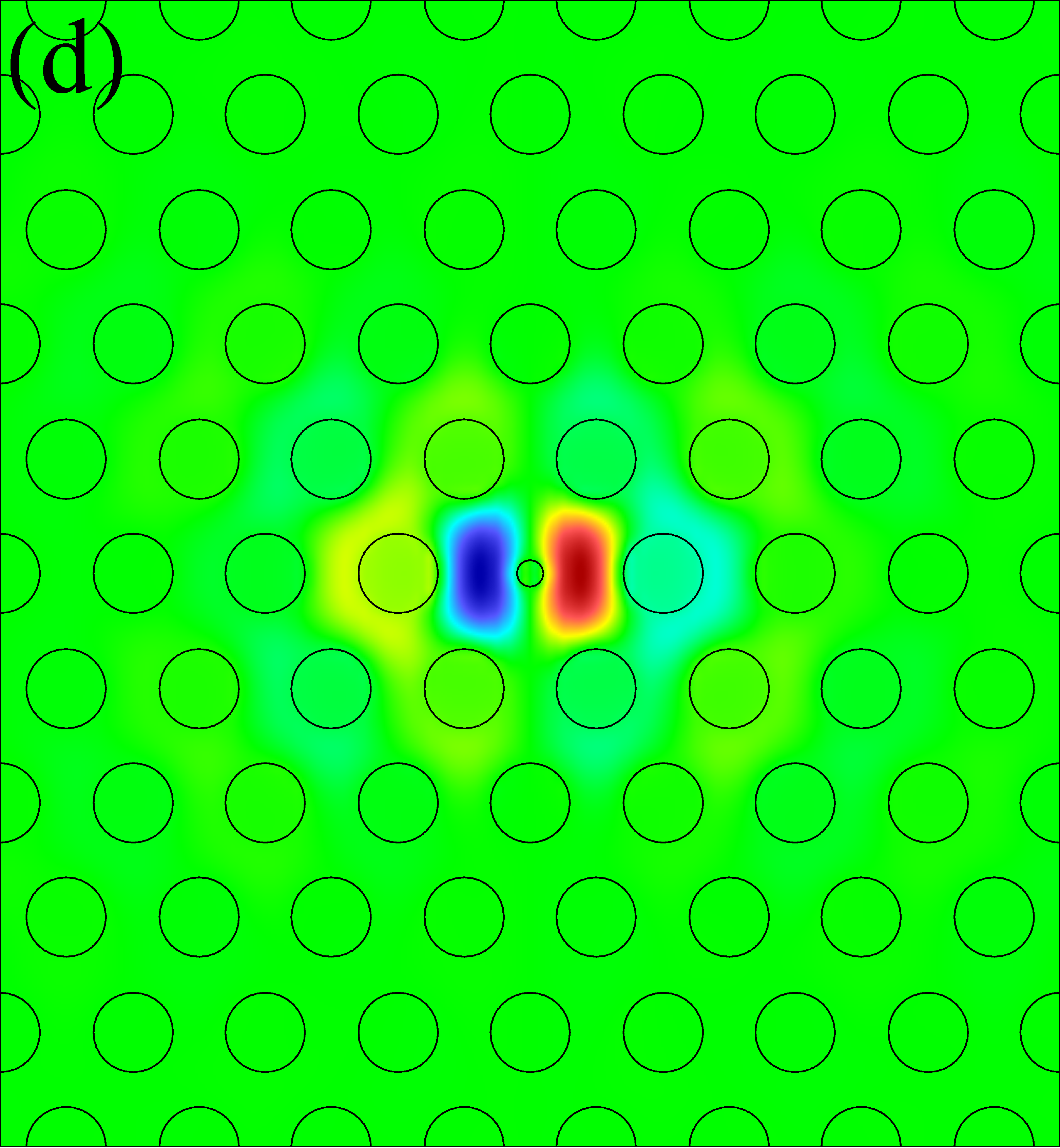} 
\includegraphics[width=0.25\linewidth]{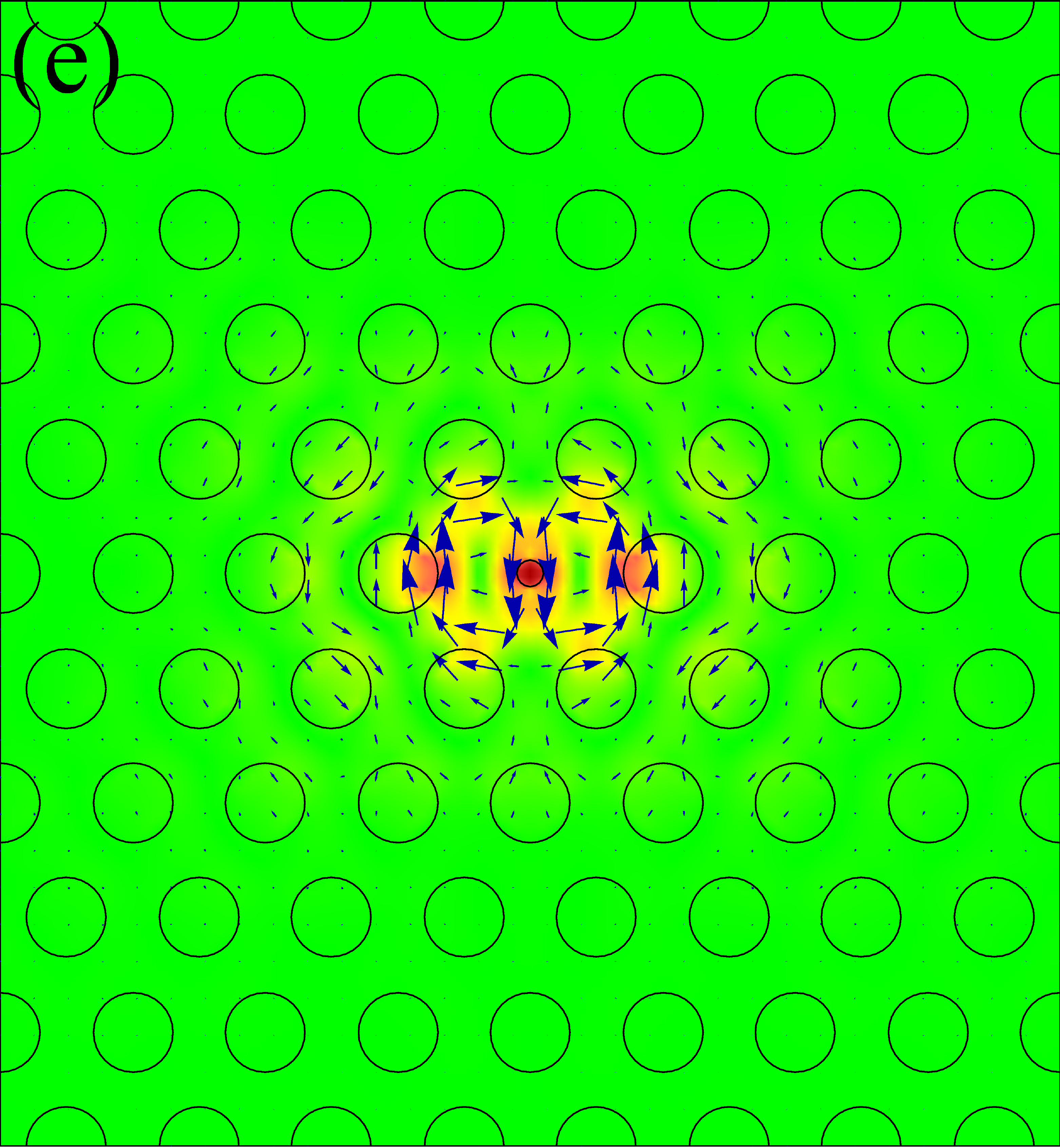} 
\includegraphics[width=0.25\linewidth]{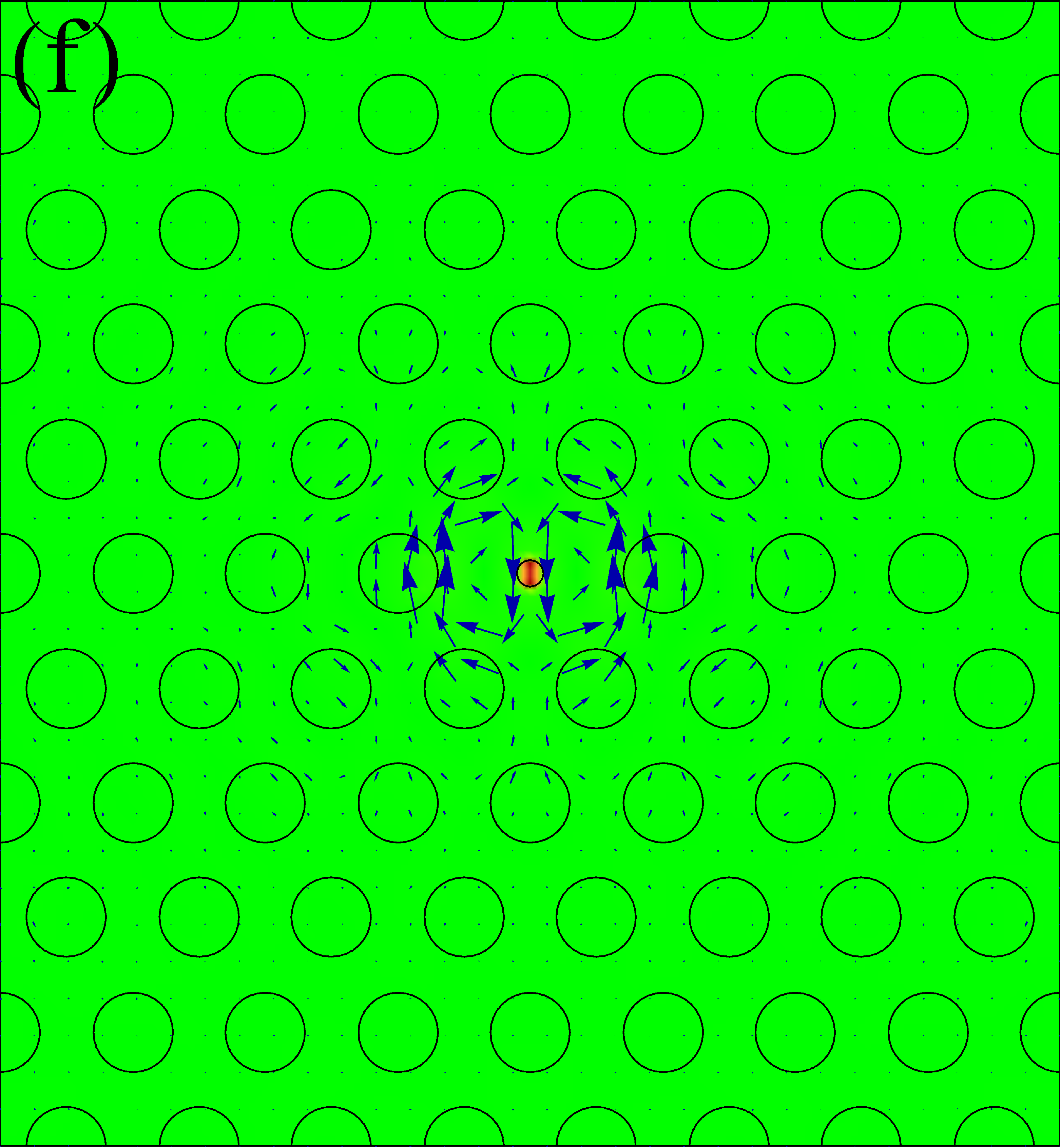} \hspace{1mm} 
\includegraphics[width=0.03\linewidth]{band} 
\caption{Field distribution of (a) $\Im\{\mathscr{H}_z\}$ of first mode, (b) $\Re\{\vec{\mathscr{E}}\}$ of first mode, (c) $\Im\{\vec{\mathscr{E}}\}$ of first mode, (d) $\Im\{\mathscr{H}_z\}$ of second mode, (e) $\Re\{\vec{\mathscr{E}}\}$ of second mode, (f) $\Im\{\vec{\mathscr{E}}\}$ of second mode for a single cavity. }
\label{f2} 
\end{figure*}

To utilize tight-binding method we should first obtain modes of a single cavity. Each of the designed  cavities has two orthogonal degenerate modes that are shown in figure \ref{f2}. 
If we represent the electric fields of these modes at resonant frequency $\nu$, by $\vec{\mathscr{E}}_{\nu,l} \ , \ \left(l=1,2\right) $, then according to Maxwell's first and second equations we can write
\begin{equation}\label{E3-1}
\nabla \times \nabla \times \vec{\mathscr{E}}_{\nu,l} \left(\mathbf{r}\right) = \varepsilon_r \left(\mathbf{r}\right) \frac{\nu^2}{c^2} \vec{\mathscr{E}}_{\nu,l} \left(\mathbf{r}\right).
\end{equation}
In this equation $\varepsilon_r \left(\mathbf{r}\right)$ is the relative permittivity profile of a single cavity and $c$ is the speed of light in vacuum.
Since eigen modes of the cavity are orthogonal we have
\begin{equation}\label{E3-2}
\int \varepsilon_r \left(\mathbf{r}\right) \vec{\mathscr{E}}_{\nu,l}^\ast \left(\mathbf{r}\right) \cdot \vec{\mathscr{E}}_{\nu,k} \left(\mathbf{r}\right) \ud \mathbf{r} = \delta_{lk}\quad , \quad l,k=1,2 ,
\end{equation}
where we have normalized the modes. According to Bloch theorem, electric field in the super crystal can be written as
\begin{eqnarray}\label{E3-3}
&\mathbf{E}_{\bm{\kappa}}\left(\mathbf{r}\right) \simeq \nonumber \\
&\sum_{n,m=-\infty}^{+\infty} \sum_{l=1}^2
 e^{-\imath \left(n \bm{\kappa}\cdot\mathbf{a}_1 + m \bm{\kappa}\cdot \mathbf{a}_2\right)} b_l \vec{ \mathscr{E}}_{\nu,l} \left(\mathbf{r}- n \mathbf{a}_1 - m \mathbf{a}_2\right) ,
\end{eqnarray}
where $\mathbf{a}_1$ and $\mathbf{a}_2$ are primitive vectors of the CPCRA lattice, and $b_1$ and $b_2$ are constants to be determined. Similar to equation (\ref{E3-1}) we can write
\begin{equation}\label{E3-4}
\nabla \times \nabla \times \mathbf{E}_{\bm{\kappa}} = \epsilon_r \left(\mathbf{r}\right) \frac{\omega^2}{c^2} \mathbf{E}_{\bm{\kappa}} ,
\end{equation}
for the whole crystal field, where $\epsilon_r \left(\mathbf{r}\right)$ is the profile of the super crystal relative permittivity and $\omega$ is the Bloch wave frequency.
Now if we replace $\mathbf{E}_{\bm{\kappa}}$ in equation (\ref{E3-4}) by its value from equation (\ref{E3-3}) we have
\begin{eqnarray}\label{E3-5}
& \sum_{n,m=-\infty}^{+\infty} \sum_{l=1}^2
\Big[ e^{-i \left(n \bm{\kappa}\cdot\mathbf{a}_1 + m \bm{\kappa}\cdot \mathbf{a}_2\right)} b_l \nonumber\\
 & \qquad \qquad \qquad \nabla\times \nabla\times \vec{ \mathscr{E}}_{\nu,l} \left(\mathbf{r}- n \mathbf{a}_1 - m \mathbf{a}_2\right) \Big]  \nonumber \\
&= \epsilon_r \left(\mathbf{r}\right) \frac{\omega^2}{c^2}
 \sum_{n,m=-\infty}^{+\infty} \sum_{l=1}^2
\Big[ e^{-\imath \left(n \bm{\kappa}\cdot\mathbf{a}_1 + m \bm{\kappa}\cdot \mathbf{a}_2\right)} b_l  \nonumber \\
 & \qquad \qquad \qquad \qquad \quad \vec{ \mathscr{E}}_{\nu,l} \left(\mathbf{r}- n \mathbf{a}_1 - m \mathbf{a}_2\right) \Big] .
\end{eqnarray}
Replacing $\nabla\times \nabla\times \vec{ \mathscr{E}}_{\nu,l} \left(\mathbf{r}- n \mathbf{a}_1 - m \mathbf{a}_2\right) $ by its value from equation (\ref{E3-1}) and inner producting both sides of equation (\ref{E3-5}) by $ \vec{ \mathscr{E}}_{\nu,k}^\ast \left(\mathbf{r}\right) $ and integrating over the entire $x-y$ plane results in
\begin{eqnarray}\label{E3-10}
&\nu^2\sum_{n,m=-\infty}^{+\infty} \sum_{l=1}^2 b_l
 e^{-\imath \left(n \bm{\kappa}\cdot\mathbf{a}_1 + m \bm{\kappa}\cdot \mathbf{a}_2\right)}\alpha_{n,m}^{k,l} \nonumber \\
&\quad =\omega^2
 \sum_{n,m=-\infty}^{+\infty} \sum_{l=1}^2 b_l
 e^{-\imath \left(n \bm{\kappa}\cdot\mathbf{a}_1 + m \bm{\kappa}\cdot \mathbf{a}_2\right)}\beta_{n,m}^{k,l} ,
\end{eqnarray}
where
\begin{eqnarray}
\alpha_{n,m}^{k,l}&=\int \vec{ \mathscr{E}}_{\nu,k}^\ast \left(\mathbf{r}\right) \cdot \vec{\mathscr{E}}_{\nu,l} \left(\mathbf{r}- n \mathbf{a}_1 - m \mathbf{a}_2\right) \nonumber\\
& \qquad \qquad \qquad \quad \varepsilon_r \left(\mathbf{r}- n \mathbf{a}_1 - m  \mathbf{a}_2\right) \ud \mathbf{r}, \nonumber \\ 
\beta_{n,m}^{k,l}&=\int \vec{ \mathscr{E}}_{\nu,k}^\ast \left(\mathbf{r}\right) \cdot  \vec{ \mathscr{E}}_{\nu,l} \left(\mathbf{r}- n \mathbf{a}_1 - m \mathbf{a}_2\right) \epsilon_r\left(\mathbf{r}\right) \ud \mathbf{r}. \nonumber
\end{eqnarray}
Using the approximations
\begin{equation*}
\alpha_{n,m}^{k,l} \simeq \beta_{n,m}^{k,l} \simeq 0 , \quad 2\leq |n| , |m|,
\end{equation*}
which is valid for confined cavity fields and doing some simplification we finally obtain the following eigenvalue problem.
\begin{equation}\label{E3-13}
\nu^2 \
\mathbb{A}\left(\bm{\kappa}\right) 
\left( \begin{array}{c}
b_1\\
b_2 
\end{array} \right) =  \omega^2 \
\mathbb{B}\left(\bm{\kappa}\right) 
\left( \begin{array} {c}
b_1 \\
b_2
\end{array} \right).
\end{equation}

\begin{figure}[ht]
\centering
\includegraphics[width=\linewidth]{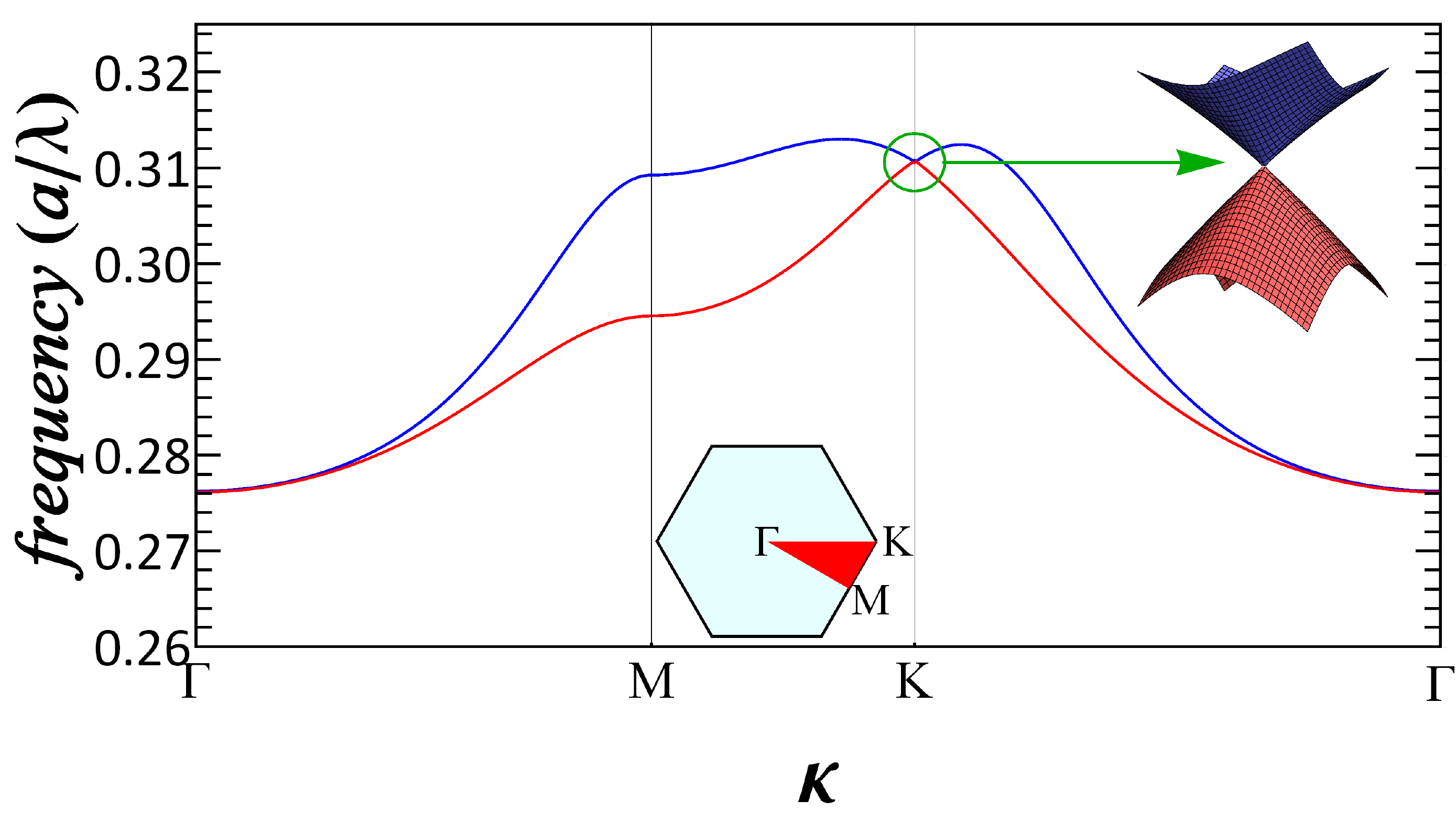}
\caption{\label{f3} The two bands of PC slab which construct the Dirac cone. They are calculated using tight-binding method}
\end{figure}

The two bands which are shown in figure \ref{f3} construct the Dirac cone and are obtained by solving this problem.

\section{System Hamiltonian}

Now we position a quantum dot inside one of the super crystal cavities. To find its behavior due to interaction with Dirac modes we have to determine the system Hamiltonian. We assume the quantum dot is at position $\mathbf{r}_0$ and an electron is inside it. The electron Hamiltonian equals
\begin{eqnarray} 
\mathbb{H}' &= \frac{1}{2m_e}(\mathbb{P} - \bar{e} \mathbb{A})^2 + \bar{e} \mathbb{V}(\mathbf{r}_0) \nonumber \\
&= \frac{\mathbb{P}^2}{2m_e} + \bar{e} \mathbb{V}(\mathbf{r}_0) - \frac{\bar{e}}{2m_e} \left( \mathbb{P} \cdot \mathbb{A} + \mathbb{A} \cdot \mathbb{P} \right) + \frac{\bar{e}^2}{2m_e} \mathbb{A}^2 \nonumber ,
\end{eqnarray}
where $m_e$, $\bar{e}$ and $\mathbb{P}$ are the electron mass, charge and momentum respectively, $\mathbb{A}$ is the magnetic vector potential,
\begin{equation*} 
\mathbb{H}_\mathrm{a} = \frac{\mathbb{P}^2}{2m_e} + \bar{e} \mathbb{V}(\mathbf{r}_0),
\end{equation*}
is the electron Hamiltonian in the absence of the field,
\begin{equation*} 
\mathbb{H}_{\mathrm{I}} = - \frac{\bar{e}}{2m_e} \left( \mathbb{P} \cdot \mathbb{A} + \mathbb{A} \cdot \mathbb{P} \right),
\end{equation*}
is the interaction of electron momentum with the field, and 
\begin{equation*}
\mathbb{H}_{\mathrm{ff}} =  \frac{\bar{e}^2}{2m_e} \mathbb{A}^2 ,
\end{equation*}
is the interaction of different field modes through coupling with electron.
Because of $\bar{e}^2$ term, $\mathbb{H}_\mathrm{ff}$ is much smaller than other Hamiltonians and can be ignored without significant error.  

We can write $\mathbb{H}_\mathrm{a}$ according to the quantum dot energy levels. If we show its energy states by 
$ \ket{i} ,  (i=1, \dots , N_s)$ we have
\begin{equation*} 
\mathbb{H}_\mathrm{a} \ket{i} = E_i \ket{i},
\end{equation*}
where $E_i$ is the energy of i'th state. Since energy eigen states create an orthonormal basis we can write
\begin{equation} \label{E4-1}
\mathbb{H}_\mathrm{a} = \mathbb{H}_\mathrm{a} \mathbb{I}  =  \mathbb{H}_\mathrm{a} \sum_i \ketbra{i}{i} = \sum_i E_i \ketbra{i}{i} = \sum_i \hbar \omega_i \hat{\sigma}_{ii} .
\end{equation}
In this relation $\omega_i = E_i / \hbar$ is the frequency related to the i'th energy level and $\hat{\sigma}_{ij} = \ketbra{i}{j}$ is the atomic ladder operator. This operator has the following properties,  
\begin{equation*} 
\hat{\sigma}_{ij} \ket{k} = \delta_{jk} \ket{i}\ , \ \hat{\sigma}_{ij}^{\dagger} = \hat{\sigma}_{ji}\ , \ \left[ \hat{\sigma}_{ij} , \hat{\sigma}_{kl} \right] = \delta_{jk} \hat{\sigma}_{il} - \delta_{li} \hat{\sigma}_{jk} ,
\end{equation*}
where $\delta_{ij}$ is the Kronecker delta function.

To write $\mathbb{H}_\mathrm{I}$ based on atomic and field operators, we have to write $\mathbb{P}$ based on atomic ladder operators. For this purpose we write
\begin{eqnarray} 
\mathbb{P} &= \mathbb{I} \ \mathbb{P} \ \mathbb{I} = \left( \sum_i \ketbra{i}{i} \right) \mathbb{P}  \Bigg( \sum_j \ketbra{j}{j} \Bigg) \nonumber \\
&= \sum_{i,j} \bra{i} \mathbb{P} \ket{j} \ketbra{i}{j} = \sum_{i,j} p_{ij} \hat{\sigma}_{ij} . \nonumber
\end{eqnarray}
Using the commutator relation 
\begin{equation*}
\left[ \mathbb{H}_\mathrm{a} , \mathbb{R} \right] \equiv \frac{\hbar}{\imath} \frac{\mathbb{P}}{m_e},
\end{equation*}
between atom Hamiltonian and position operator, $\mathbb{R}$, we can write
\begin{eqnarray} \label{E4-2}
p_{ij} &= \bra{i} \mathbb{P} \ket{j} = \imath  \frac{m_e}{\hbar} \bra{i} \left[ \mathbb{H}_\mathrm{a} , \mathbb{R} \right] \ket{j} \nonumber \\
&= \imath  \frac{m_e}{\hbar} \bra{i} \left( \sum_k \hbar \omega_k \hat{\sigma}_{kk} \mathbb{R} - \mathbb{R} \sum_k \hbar \omega_k \hat{\sigma}_{kk} \right)  \ket{j} \nonumber \\
&= \imath  m_e \left(\omega_i - \omega_j \right) \bra{i} \mathbb{R} \ket{j} = \imath \frac{m_e}{\bar{e}} \omega_{ij} \bra{i} \bar{e} \mathbb{R} \ket{j} \nonumber \\
&= \imath \frac{m_e}{\bar{e}} \omega_{ij} \mathbf{d}_{ij},
\end{eqnarray} 
where $\omega_{ij}= \omega_i - \omega_j$ is the transition frequency between i'th and j'th energy levels and $ \mathbf{d}_{ij} = \bra{i} \bar{e} \mathbb{R} \ket{j} $ is the transition electric dipole between $\ket{i}$ and $\ket{j}$ states. 
Here we have replaced the value of $\mathbb{H}_\mathrm{a}$ from equation (\ref{E4-1}).

Now we should write magnetic vector potential operator based on photon annihilation, $\hat{a}$, and creation, $\hat{a}^{\dagger}$, operators. 
If there is only one cavity in the basis PC, then $\mathbb{A}$ can be written as \cite{Schleich4}
\begin{equation*} 
\mathbb{A}(\mathbf{r},t) = \sum_{\ell} \sqrt{\frac{\hbar}{2 \epsilon \Omega_{\ell}}} \left[ \bm{u}_{\ell} (\mathbf{r}) \hat{a}_{\ell}(t) + \bm{u}^*_{\ell} (\mathbf{r}) \hat{a}^{\dagger}_{\ell}(t) \right] , 
\end{equation*}
where the sum is over all the cavity modes, $\Omega_{\ell}$ is the frequency of the $\ell$'th mode, $\bm{u}_{\ell}(\mathbf{r})$ is the normalized $\ell$'th mode function, $\epsilon$ is the electric permittivity of space, and $\hbar$ is the reduced Planck constant.
We know mode functions are orthonormal, that is
\begin{equation*}
\int \bm{u}_{\ell} (\mathbf{r}) \bm{u}^*_{\ell'} (\mathbf{r}) \ud \mathbf{r} = \delta_{\ell \ell'}.
\end{equation*}

In most of quantum optics texts, $|\bm{u}_{\ell} (\mathbf{r})|^2_\mathrm{max}$ is shown by $1/\mathcal{V}_{\ell}$ \cite{Schleich4,Claudio4}, where $\mathcal{V}_{\ell}$ is called the effective volume of $\ell$'th mode.
So we can write vector potential operator as
\begin{equation} \label{E4-3}
\mathbb{A}(\mathbf{r},t) = \sum_{\ell} \sqrt{\frac{\hbar}{2 \epsilon \Omega_{\ell} \mathcal{V}_{\ell} }} \left[ \bm{v}_{\ell} (\mathbf{r}) \hat{a}_{\ell}(t)  + \bm{v}^*_{\ell} (\mathbf{r}) \hat{a}^{\dagger}_{\ell}(t)  \right],
\end{equation}
where $\bm{v}_{\ell} (\mathbf{r})$ is the mode function with its maximum limited to one.
If there is more than one, say $N_c$, cavity in the basis PC, then $\mathbb{A}$ becomes
\begin{equation} \label{E4-4}
\mathbb{A} = \sum_{\ell} \sqrt{\frac{\hbar}{2 \epsilon \Omega_{\ell} N_c \mathcal{V}_{\ell} }} \left[ \bm{v}_{\ell} (\mathbf{r}) \hat{a}_{\ell}(t)  + \bm{v}^*_{\ell} (\mathbf{r}) \hat{a}^{\dagger}_{\ell}(t)  \right].
\end{equation}
Note that here we have obviously more modes than in the case of a single cavity which has only two modes.
Therefore $\mathbb{H}_\mathrm{I}$ becomes
\begin{eqnarray} \label{E4-7}
\mathbb{H}_\mathrm{I} &= - \frac{\bar{e}}{2m_e} \left( \mathbb{P} \cdot \mathbb{A} + \mathbb{A} \cdot \mathbb{P} \right) = - \frac{\bar{e}}{m_e} \mathbb{A} \cdot \mathbb{P} \nonumber \\
& = - \frac{\bar{e}}{m_e} \sum_{\ell} \sum_{i,j} \sqrt{\frac{\hbar}{2 \epsilon \Omega_{\ell} N_c \mathcal{V}_{\ell} }} \left[ \bm{v}_{\ell} (\mathbf{r}_0) \hat{a}_{\ell} + \bm{v}^*_{\ell} (\mathbf{r}_0) \hat{a}^{\dagger}_{\ell} \right] \nonumber \\
& \qquad \cdot \left( \imath \frac{m_e}{\bar{e}} \omega_{ij} \mathbf{d}_{ij} \hat{\sigma}_{ij} \right)  \nonumber \\
& = -\imath \hbar \sum_{\ell} \sum_{i,j} \left[ g_{ij \ell} (\mathbf{r}_0) \hat{a}_{\ell} \hat{\sigma}_{ij} - g_{ij \ell}^* (\mathbf{r}_0) \hat{a}^{\dagger}_{\ell} \hat{\sigma}^{\dagger}_{ij} \right] ,
\end{eqnarray}
where 
\begin{equation} \label{E4-8}
g_{ij \ell} (\mathbf{r}) = \frac{\omega_{ij}}{\hbar} \sqrt{\frac{\hbar}{2 \epsilon \Omega_{\ell} N_c \mathcal{V}_{\ell} }}  \bm{v}_{\ell} (\mathbf{r}) \cdot \mathbf{d}_{ij} ,
\end{equation}
is the Rabi frequency.
In equation (\ref{E4-7}) we have used the fact that atomic and photonic operators commute with each other, that is 
\begin{equation*}
\left[\hat{a}_{\ell} , \hat{\sigma}_{ij} \right]=0 \quad , \quad [\hat{a}^{\dagger}_{\ell} , \hat{\sigma}_{ij} ]=0 \quad \forall i,j,\ell.
\end{equation*}
In rotating wave approximation (RWA), $\mathbb{H}_\mathrm{I}$ simplifies to 
\begin{equation} \label{E4-9}
\mathbb{H}_\mathrm{I}  = -\imath \hbar \sum_{\ell} \sum_{i>j} \left[ g_{ij \ell} (\mathbf{r}_0) \hat{a}_{\ell} \hat{\sigma}_{ij} - g_{ij \ell}^* (\mathbf{r}_0) \hat{a}^{\dagger}_{\ell} \hat{\sigma}^{\dagger}_{ij} \right] ,
\end{equation}

Finally Hamiltonian of the whole system becomes 
\begin{equation*}
\mathbb{H}_t = \mathbb{H}' + \mathbb{H}_\mathrm{f},
\end{equation*}
where 
\begin{equation*}
\mathbb{H}_\mathrm{f} = \int \left[ \frac{\epsilon}{2} \mathbb{E}^2 (\mathbf{r}) + \frac{1}{2 \mu} \mathbb{B}^2 (\mathbf{r}) \right] \ud \mathbf{r} ,
\end{equation*}
is the Hamiltonian of the electromagnetic field and $\mathbb{E}$ and $\mathbb{B}$ are the electric field and magnetic field operators respectively. By writing $\mathbb{E}$ and $\mathbb{B}$ in terms of $\mathbb{A}$ and using equation (\ref{E4-3}), $\mathbb{H}_\mathrm{f}$ can be written as 
\begin{equation*}
\mathbb{H}_\mathrm{f} = \sum_{\ell} \hbar \Omega_{\ell} \left( \hat{a}^{\dagger}_{\ell} \hat{a}_{\ell} + \frac{1}{2} \right) .
\end{equation*}

Our system, in its simplest form, consists of a quantum dot with two energy eigen states inside one of the $N_c$ coupled cavities which their modes construct a Dirac cone. Hamiltonian of this system, considering RWA, becomes 
\begin{eqnarray} \label{E4-50}
\mathbb{H} &=  \left( \hbar \omega_1 \hat{\sigma}_{11} + \hbar \omega_2 \hat{\sigma}_{22} \right) + \sum_{\bm{\kappa},p} \hbar \Omega_{\bm{\kappa}p} \hat{a}^{\dagger}_{\bm{\kappa}p} \hat{a}_{\bm{\kappa}p} \nonumber \\
&   - \imath \hbar \sum_{\bm{\kappa},p} \left[ g_{21 \bm{\kappa}p} (\mathbf{r}_0) \hat{a}_{\bm{\kappa}p} \hat{\sigma}_{21} - g_{21 \bm{\kappa}p}^* (\mathbf{r}_0) \hat{a}^{\dagger}_{\bm{\kappa}p} \hat{\sigma}_{12} \right] , 
\end{eqnarray}
where the sums are over all the confined Dirac modes, that is the modes which fall below the light cone of the super crystal. In this equation $p=1$ and $p=2$ denote the lower and upper parts of the cone respectively.
We have neglected zero point energy of the field which only shifts all the system energy levels by a constant value.
We have also assumed the interaction with other modes of the crystal is not very different from  interaction with vacuum and have omitted it in this equation.

In the remaining, without losing the problem generality, we set $\mathbf{r}_0 = 0$.
 As said, a single cavity has two orthogonal modes that are shown in figure \ref{f2}. The electric fields of these two modes are orthogonal at the center of the cavity too. We now assume the quantum dot transition dipole is parallel to the first and perpendicular to the second. Hence considering equation (\ref{E3-3}), we conclude $\sqrt{1/\mathcal{V}_{\bm{\kappa}p}}$ is approximately $b_{1,p}(\bm{\kappa})$ times of the case with just a single cavity. Again $p$ marks lower and upper bands constructing Dirac cone.

\section{Atom evolution}
\begin{figure*}[!ht]
\centering
\includegraphics[width=0.35\textwidth]{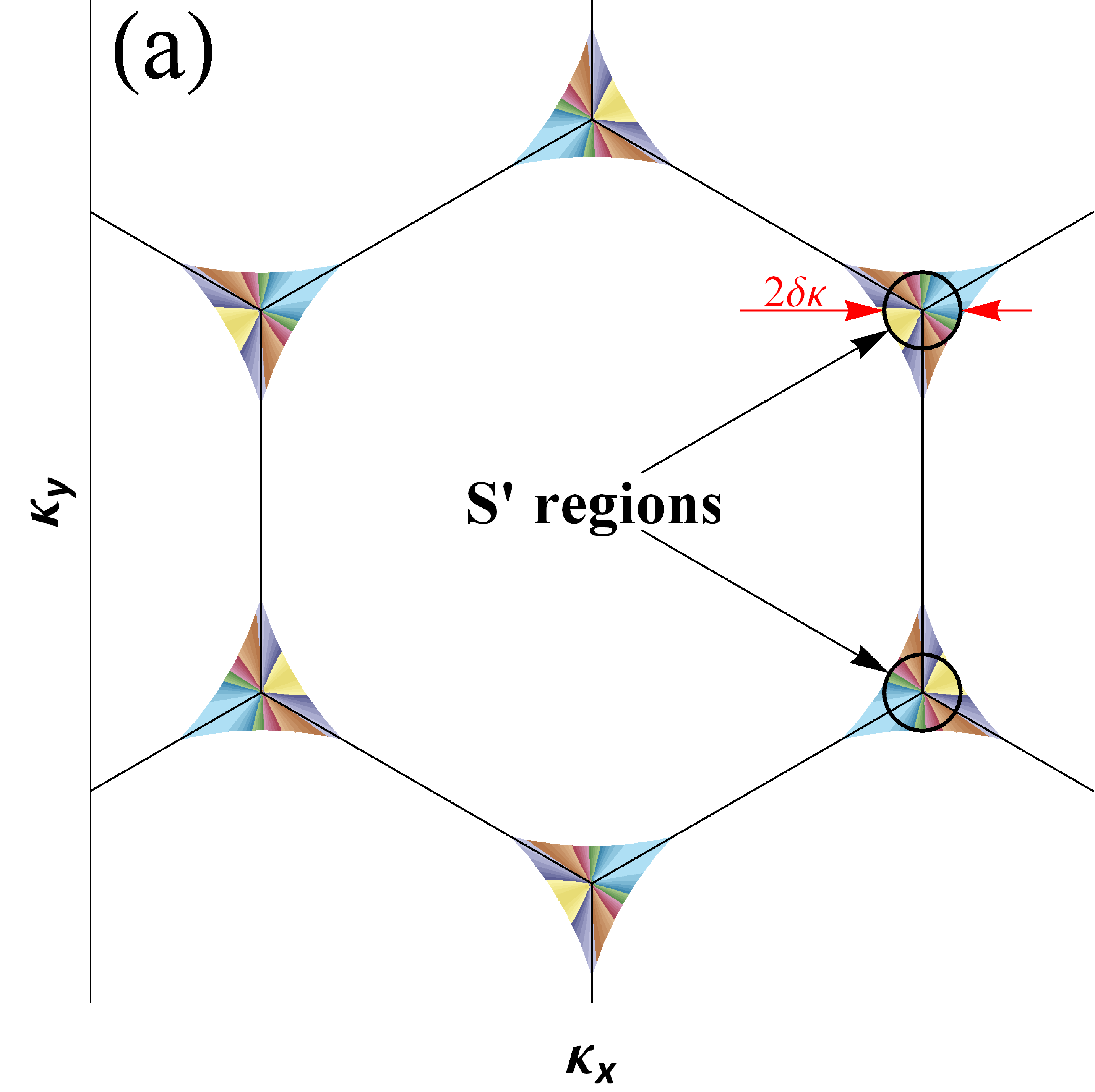} \hspace{1 cm}
\includegraphics[width=0.35\textwidth]{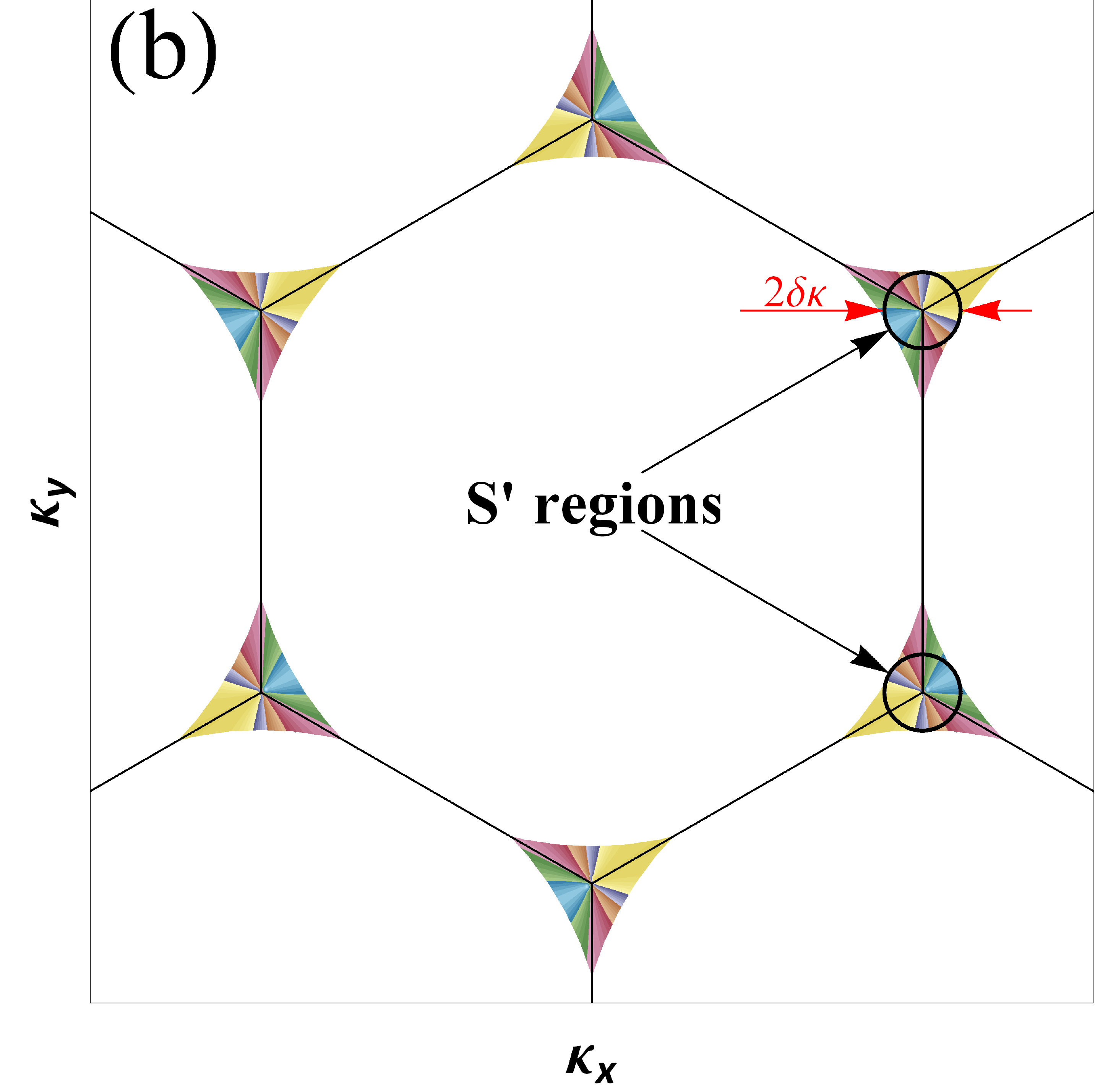} \hspace{1 cm}
\includegraphics[width=0.057\textwidth]{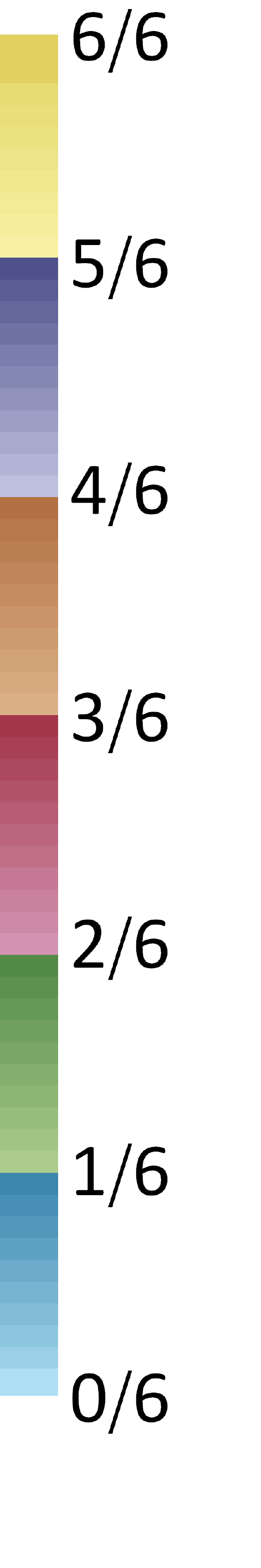}\\
\hspace{-1.5 cm}
\includegraphics[width=0.25\textwidth]{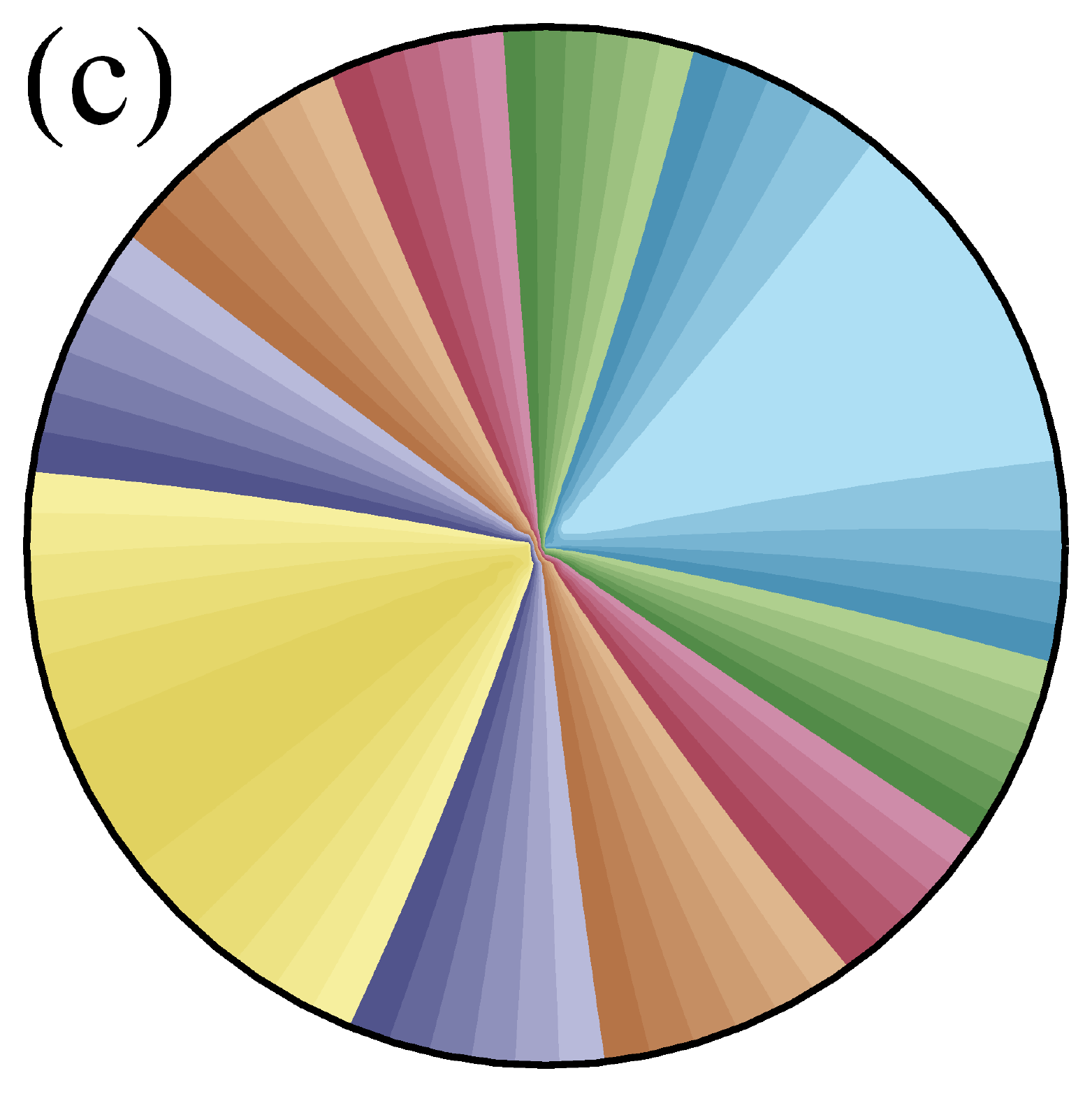} \hspace{3 cm}
\includegraphics[width=0.25\textwidth]{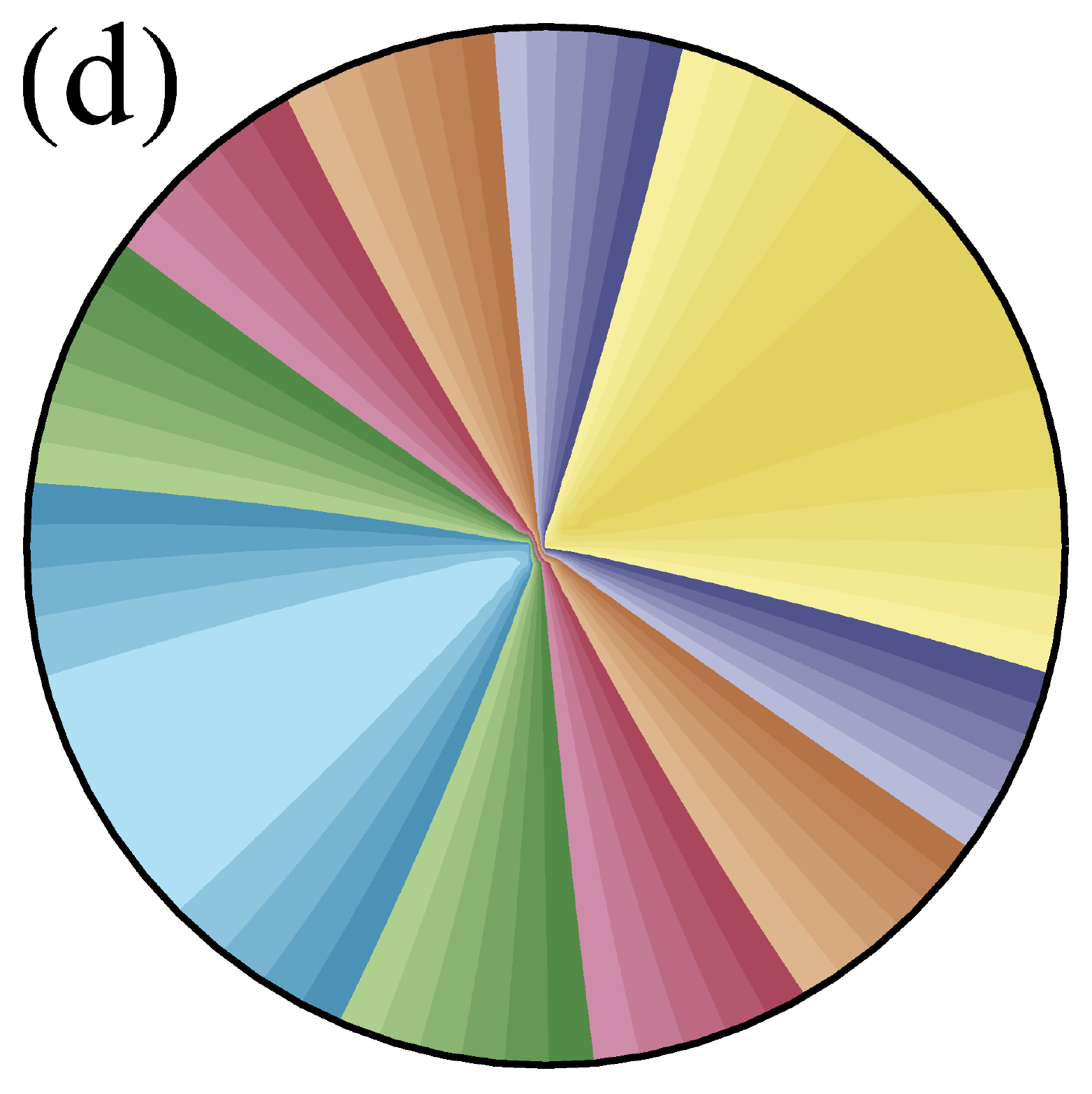} \hfill
\caption{Variation of (a) $|b_{1,1}(\bm{\kappa})|^2$ and (b) $|b_{1,2}(\bm{\kappa})|^2$ as constant contours in reciprocal lattice and in regions below the light cone. $S'$ regions are shown as two circles in both figures. The upper $S'$ region in each of them is plotted after zooming in figures (c) and (d)}
\label{f4} 
\end{figure*}

\begin{figure}[ht]
\centering
\includegraphics[width=\linewidth]{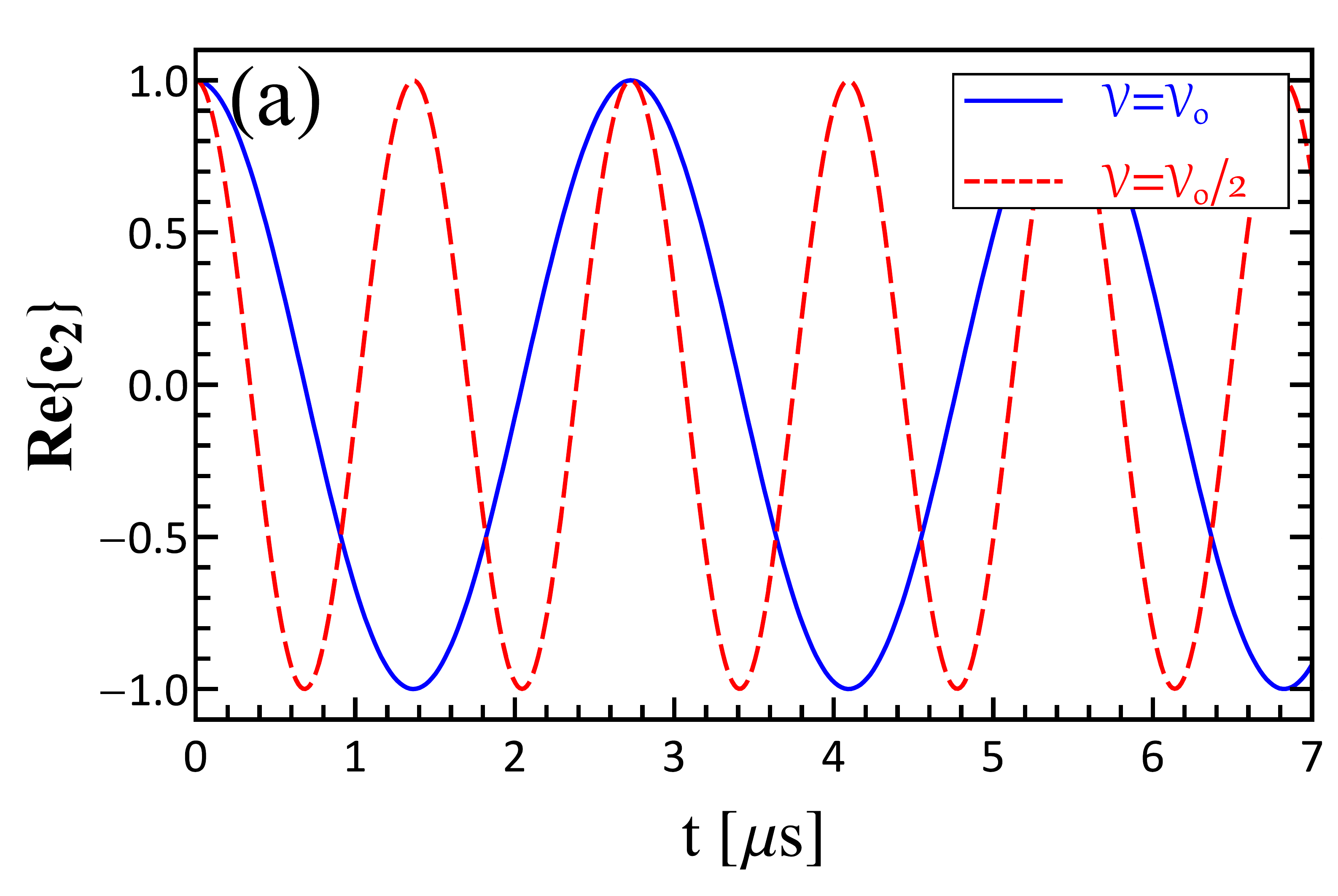} \\
\includegraphics[width=\linewidth]{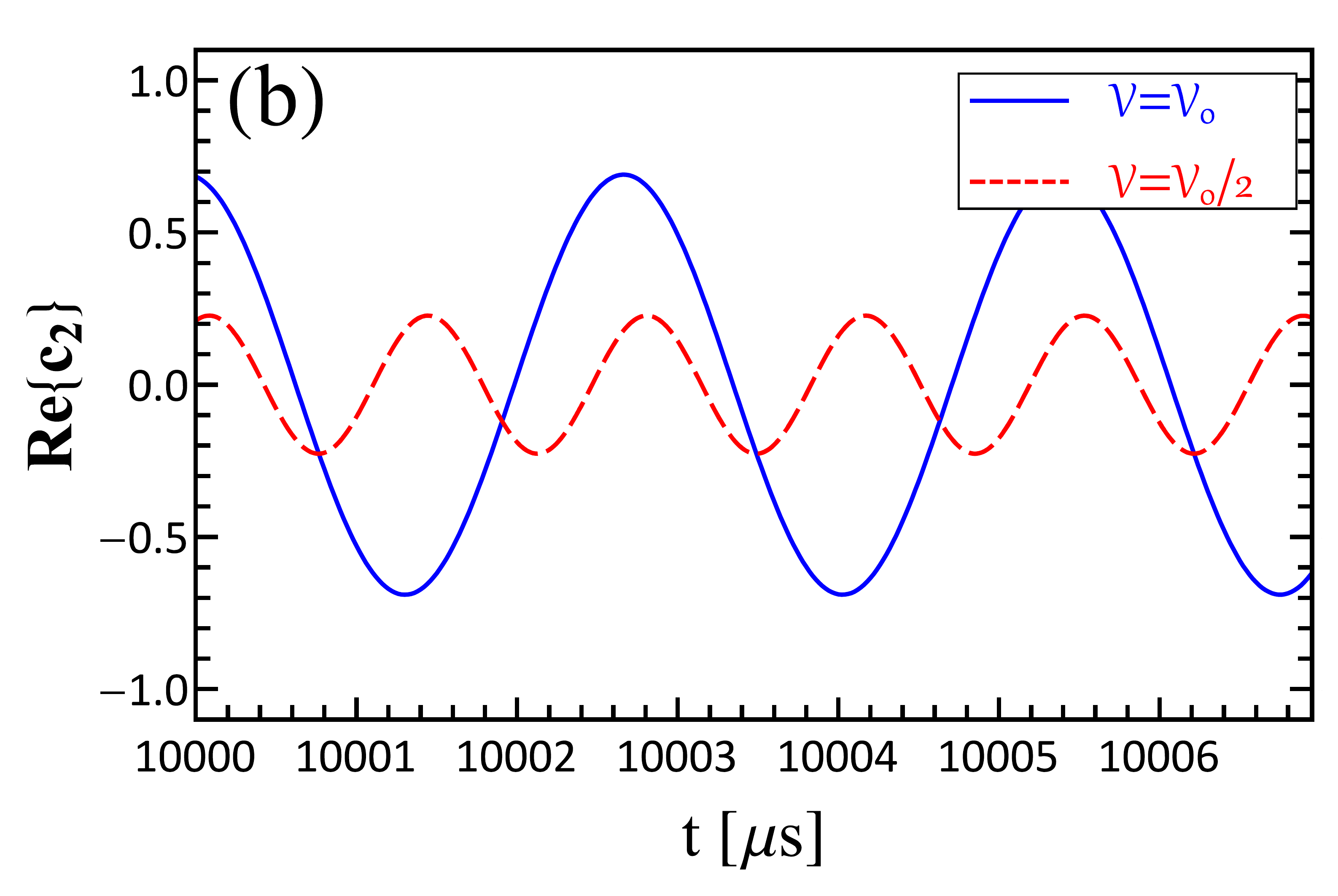}
\caption{\label{f5} $\Re\{c_2(t)\}$ with $d_{21}=100 \ \mathrm{Debye}$, $\lambda_{21} =1.55 \mu \mathrm{m}$, $\Omega_\mathrm{D}= \omega_{21}$, $\delta \kappa = \Gamma \mathrm{K} /10$, $\alpha = 5.38 \times 10^7 \mathrm{m} / \mathrm{s}$, $N_c=7$, $\varphi_\mathrm{a}=0^{\circ}$, $\theta_\mathrm{a}=90^{\circ}$, for two cavity mode volumes and two time intervals.}
\end{figure}

For simplicity we assume there is initially no photon in the system and the atom is in a superposition state of its ground and exited states. So we can write the initial state of the system as
\begin{eqnarray} \label{E4-19}
\ket{\Psi(t=0)} &= \ket{\psi_\mathrm{atom}} \otimes \ket{\psi_\mathrm{field}} \nonumber \\
&= c_1(0) \ket{1,\{0\}} + c_2(0) \ket{2,\{0\}} ,
\end{eqnarray}
where $c_1(0)$ and $c_2(0)$ can be written in the most general case as \cite{Nielsen4}
\begin{equation} \label{E4-14}
c_1(0) = \cos \theta_\mathrm{a} \quad , \quad c_2(0) = e^{\imath \varphi_\mathrm{a}} \sin \theta_\mathrm{a} .
\end{equation}
The state of the system in times $t>0$ can be written as
\begin{eqnarray} \label{E4-52}
\ket{\Psi(t)} &= c_1(t) e^{- \imath \omega_1 t} \ket{1,\{0\}} + c_2(t) e^{- \imath \omega_2 t} \ket{2,\{0\}} \nonumber \\
&+  \sum_{\bm{\kappa},p} c_{\bm{\kappa}p}(t) e^{- \imath (\omega_1+ \Omega_{\bm{\kappa}p}) t} \ket{1,\{1_{\bm{\kappa}p}\}}  ,
\end{eqnarray}
where the effect of
\begin{equation*}
\mathbb{H}_0 = \left( \hbar \omega_1 \hat{\sigma}_{11} + \hbar \omega_2 \hat{\sigma}_{22} \right) + \sum_{\bm{\kappa},p} \hbar \Omega_{\bm{\kappa}p} \hat{a}^{\dagger}_{\bm{\kappa}p} \hat{a}_{\bm{\kappa}p}  ,
\end{equation*} 
has been included in the state via exponential terms.
By inserting the system ket state from equation (\ref{E4-52}) and the system Hamiltonian from 
equation (\ref{E4-50}) into Schr\"{o}dinger equation
\begin{equation*}
\imath \hbar \frac{\partial }{\partial t}\ket{\Psi(t)} = \mathbb{H} \ket{\Psi (t)} ,
\end{equation*}
we reach
\begin{eqnarray} 
&\imath \hbar \bigg[ \dot{c}_1(t) e^{- \imath \omega_1 t} \ket{1,\{0\}} +  \dot{c}_2(t) e^{- \imath \omega_2 t} \ket{2,\{0\}} \nonumber\\
& \qquad \qquad \quad +  \sum_{\bm{\kappa},p} \dot{c}_{\bm{\kappa}}(t) e^{- \imath (\omega_1+ \Omega_{\bm{\kappa}p}) t} \ket{1,\{1_{\bm{\kappa}p}\}} \bigg] \nonumber \\
& =  \imath \hbar c_2(t) e^{- \imath \omega_2 t} \sum_{\bm{\kappa},p} g_{21 \bm{\kappa}p}^* (\mathbf{r}_0) \ket{1,\{1_{\bm{\kappa}p}\}} \nonumber\\
& \qquad \quad - \imath \hbar  \sum_{\bm{\kappa},p} c_{\bm{\kappa}p}(t) e^{- \imath (\omega_1+ \Omega_{\bm{\kappa}p}) t} g_{21 \bm{\kappa}p} (\mathbf{r}_0) \ket{2,\{0\}} . \nonumber
\end{eqnarray}
By equating the coefficients of similar kets at both sides, we obtain the following system of equations.
\begin{eqnarray}
 \dot{c}_1(t) &= 0,  \nonumber \\
 \dot{c}_2(t) &= -\sum_{\bm{\kappa},p} c_{\bm{\kappa}p}(t) g_{21 \bm{\kappa}p} (\mathbf{r}_0) e^{- \imath \Delta_{\bm{\kappa}p} t} , \nonumber \\
\dot{c}_{\bm{\kappa}p}(t) &=  c_2(t) g_{21 \bm{\kappa}p}^* (\mathbf{r}_0)  e^{\imath \Delta_{\bm{\kappa}p} t}.  \nonumber 
\end{eqnarray}
The first equation indicates $c_1(t)$ do not change with time. To solve the next two equations we use Laplace transform. In Laplace domain these equations become
\numparts
\begin{eqnarray}
&s C_2(s) - c_2(0) = -\sum_{\bm{\kappa},p} C_{\bm{\kappa}p}(s+\imath \Delta_{\bm{\kappa}p}) g_{21 \bm{\kappa}p} (0) , \label{E4-53a} \\
&s C_{\bm{\kappa}p}(s) - c_{\bm{\kappa}p}(0) =  C_2(s- \imath \Delta_{\bm{\kappa}p}) g_{21 \bm{\kappa}p}^* (0)  . \label{E4-53b} 
\end{eqnarray}
\endnumparts
By replacing $C_{\bm{\kappa}p}(s+\imath \Delta_{\bm{\kappa}p})$ from equation (\ref{E4-53b}) into equation (\ref{E4-53a}) we have
\begin{equation} \label{E4-54}
C_2(s) = \frac{c_2(0)}{s+ \sum_{\bm{\kappa},p} |g_{21 \bm{\kappa}p} (0)|^2 /(s+\imath \Delta_{\bm{\kappa}p}) }.
\end{equation}
To calculate $C_2(s)$, we should first calculate $\sum_{\bm{\kappa},p} |g_{21 \bm{\kappa}p} (0)|^2 /(s+\imath \Delta_{\bm{\kappa}p}) $. Since $\bm{\kappa}$ is a continuous variable we replace the sum by an integral as
\begin{equation*} 
\sum_{\bm{\kappa},p} \rightarrow \frac{\mathcal{S}}{(2\pi)^2} \sum_{p=1}^2 \int_{S'}  \ud \bm{\kappa} ,
\end{equation*}
where $\mathcal{S}$ is the area of the super crystal unit cell and $S'$ is a region in reciprocal lattice which besides being in the first Brillouin zone, is on the Dirac cone and below the light cone.
Since Dirac point is located on the high symmetry point $\mathrm{K}$, at the corners of the Brillouin zone, $S'$ regions can be approximated by two circles centered at $\mathrm{K}$ point and with radius $\delta \kappa$. This radius is a function of Dirac point frequency and Dirac cone shape.
For the lower part of Dirac cone ($p=1$), we have
\begin{eqnarray} \label{E4-55}
&\sum_{\bm{\kappa}} \frac{|g_{21 \bm{\kappa}1} (0)|^2}{(s+\imath \Delta_{\bm{\kappa}1})} \\
&= \frac{\mathcal{S}}{(2\pi)^2} \int_{S'} \frac{\omega_{21}^2}{\hbar} \frac{|b_{1,1}(\bm{\kappa})|^2}{2 \epsilon_0 \Omega_{\bm{\kappa}1} N_c \mathcal{V} }  \frac{d_{21}^2}{(s+\imath \Delta_{\bm{\kappa}1})} \ud \bm{\kappa} \nonumber \\
&= \frac{\mathcal{S}\omega_{21}^2 d_{21}^2}{2(2\pi)^2 \hbar  \epsilon_0 N_c \mathcal{V} } \int_{S'}  \frac{|b_{1,1}(\bm{\kappa})|^2}{\Omega_{\bm{\kappa}1} }  \frac{1}{\left[ s+\imath (\Omega_{\bm{\kappa}1}- \omega_{21}) \right]} \ud \bm{\kappa} \nonumber \\
&= \frac{\mathcal{S}\omega_{21}^2 d_{21}^2}{2(2\pi)^2 \hbar  \epsilon_0 N_c \mathcal{V} } \int_{S'}  \frac{|b_{1,1}(\bm{\kappa})|^2}{(\Omega_\mathrm{D}- \alpha \kappa)\left[ s+\imath (\Omega_{\bm{\kappa}1}- \omega_{21}) \right]}   \ud \bm{\kappa} \nonumber \\
&= \frac{\mathcal{S}\omega_{21}^2 d_{21}^2}{2(2\pi)^2 \hbar  \epsilon_0 N_c \mathcal{V} } \int_{S'}  \frac{|b_{1,1}(\bm{\kappa})|^2}{(\Omega_\mathrm{D}- \alpha \kappa)\left[ s+\imath (\Delta_\mathrm{D}- \alpha \kappa) \right]}   \ud \bm{\kappa} \nonumber .
\end{eqnarray}
In this equation $\Omega_\mathrm{D}$ is the frequency of Dirac point, $\Delta_\mathrm{D} =\Omega_\mathrm{D} - \omega_{21}$, $\kappa$ is the distance from $\mathrm{K}$ point and $\alpha$ is the gradient of Dirac cone.
Figure \ref{f4} shows the variation of $|b_{1,1}(\bm{\kappa})|^2$ and $|b_{1,2}(\bm{\kappa})|^2$ as constant contours in reciprocal lattice and in the regions below the light cone.
Zooming in $S'$ regions, we can see $|b_{1,1}(\bm{\kappa})|^2$ and $|b_{1,2}(\bm{\kappa})|^2$
depend mainly on azimuthal component ($\varphi$), and a little on radial component ($r$). 
Therefore equation (\ref{E4-55}) simplifies to 
\begin{eqnarray} \label{E4-56}
&\sum_{\bm{\kappa}} \frac{|g_{21 \bm{\kappa}1} (0)|^2}{(s+\imath \Delta_{\bm{\kappa}1})} \nonumber \\
&= \frac{\mathcal{S}\omega_{21}^2 d_{21}^2}{2(2\pi)^2 \hbar  \epsilon_0 N_c \mathcal{V} } \int_{S'}  \frac{|b_{1,1}(\bm{\kappa})|^2}{(\Omega_\mathrm{D}- \alpha \kappa)\left[ s+\imath (\Delta_\mathrm{D}- \alpha \kappa) \right]}   \ud \bm{\kappa} \nonumber \\
&= \frac{2\sqrt{3}a^2\omega_{21}^2 d_{21}^2}{(2\pi)^2 \hbar  \epsilon_0 N_c \mathcal{V} } \int_{S'}  \frac{|b_{1,1}(\bm{\kappa})|^2}{(\Omega_\mathrm{D}- \alpha \kappa)\left[ s-\imath  \alpha \kappa \right]}   \ud \bm{\kappa} \nonumber \\
&= \frac{2\sqrt{3}a^2\omega_{21}^2 d_{21}^2}{(2\pi)^2 \hbar  \epsilon_0 N_c \mathcal{V} } \int_0^{2\pi} |b_{1,1}(\varphi)|^2 \ud \varphi  \nonumber \\
& \qquad \qquad \qquad \cdot \int_0^{\delta \kappa} \frac{\kappa }{(\Omega_\mathrm{D}- \alpha \kappa)\left[ s-\imath  \alpha \kappa \right]}   \ud \kappa \nonumber \\
&= \frac{\sqrt{3}a^2\omega_{21}^2 d_{21}^2}{4 \pi \hbar  \epsilon_0 N_c \mathcal{V} }  \int_0^{\delta \kappa} \frac{\kappa }{(\Omega_\mathrm{D}- \alpha \kappa)\left[ s-\imath  \alpha \kappa \right]} \ud \kappa \nonumber \\
&= \frac{\sqrt{3}a^2\omega_{21}^2 d_{21}^2}{4 \pi \hbar  \epsilon_0 N_c \mathcal{V} } \cdot \frac{1}{2 \alpha^2 (s - \imath \Omega_\mathrm{D})} \bigg\{ 2 \Omega_\mathrm{D} \ln \left( \frac{\Omega_\mathrm{D}}{\Omega_\mathrm{D} - \alpha  \delta \kappa} \right)   \nonumber \\
& + s \left[ -2 \arctan \left( \frac{\alpha \delta \kappa}{s} \right) + \imath \ln \left( \frac{s^2}{s^2 + \alpha^2 \delta \kappa^2} \right) \right] \bigg \} ,
\end{eqnarray}
where we have used $\mathcal{S}=2\sqrt{3}a^2$ and 
\begin{equation*} 
\int_0^{2\pi} |b_{1,1}(\varphi)|^2 \ud \varphi \simeq \int_0^{2\pi} |b_{1,2}(\varphi)|^2 \ud \varphi \simeq \pi  ,
\end{equation*}
relations. We have also set $\Delta_\mathrm{D}=0$ for simplicity.
If we go through the same process for the upper part of Dirac cone ($p=2$), we obtain
\begin{eqnarray} \label{E4-57}
&\sum_{\bm{\kappa}} \frac{|g_{21 \bm{\kappa}2} (0)|^2}{(s+\imath \Delta_{\bm{\kappa}2})} \nonumber \\
&= \frac{2\sqrt{3}a^2\omega_{21}^2 d_{21}^2}{(2\pi)^2 \hbar  \epsilon_0 N_c \mathcal{V} } \int_0^{2\pi} |b_{1,2}(\varphi)|^2 \ud \varphi  \nonumber \\
& \qquad \qquad \qquad \cdot \int_0^{\delta \kappa} \frac{\kappa }{(\Omega_\mathrm{D}+ \alpha \kappa)\left[ s+\imath  \alpha \kappa \right]}   \ud \kappa \nonumber \\
&= \frac{\sqrt{3}a^2\omega_{21}^2 d_{21}^2}{4 \pi \hbar  \epsilon_0 N_c \mathcal{V} }  \int_0^{\delta \kappa} \frac{\kappa }{(\Omega_\mathrm{D}+ \alpha \kappa)\left[ s+\imath  \alpha \kappa \right]} \ud \kappa \nonumber \\
&= \frac{\sqrt{3}a^2\omega_{21}^2 d_{21}^2}{4 \pi \hbar  \epsilon_0 N_c \mathcal{V} } \cdot \frac{1}{2 \alpha^2 (s - \imath \Omega_\mathrm{D})} \bigg\{ 2 \Omega_\mathrm{D} \ln \left( \frac{\Omega_\mathrm{D}}{\Omega_\mathrm{D} + \alpha  \delta \kappa} \right)   \nonumber \\
& + s \left[ 2 \arctan \left( \frac{\alpha \delta \kappa}{s} \right) + \imath \ln \left( \frac{s^2}{s^2 + \alpha^2 \delta \kappa^2} \right) \right] \bigg \} .
\end{eqnarray}
Using equations (\ref{E4-56}) and (\ref{E4-57}) we get
\begin{eqnarray} 
&\sum_{\bm{\kappa},p} \frac{|g_{21 \bm{\kappa}p} (0)|^2}{(s+\imath \Delta_{\bm{\kappa}p})} \nonumber \\
&= \frac{\sqrt{3}a^2\omega_{21}^2 d_{21}^2}{4 \pi \hbar  \epsilon_0 N_c \mathcal{V} } 
 \frac{ \Omega_\mathrm{D} \ln \left( \frac{\Omega_\mathrm{D}^2}{\Omega_\mathrm{D}^2 - \alpha^2  \delta \kappa^2} \right) +
 \imath s \ln \left( \frac{s^2}{s^2 + \alpha^2 \delta \kappa^2} \right)   }{ \alpha^2 (s - \imath \Omega_\mathrm{D}) } , \nonumber
\end{eqnarray}
which results in 
\begin{eqnarray}
C_2(s) &= c_2(0) \Bigg[ s \nonumber \\
&+ \chi_{21} \frac{ \Omega_\mathrm{D} \ln \left( \frac{\Omega_\mathrm{D}^2}{\Omega_\mathrm{D}^2 - \alpha^2  \delta \kappa^2} \right) +
 \imath s \ln \left( \frac{s^2}{s^2 + \alpha^2 \delta \kappa^2} \right) }{ \alpha^2 (s - \imath \Omega_\mathrm{D}) } \Bigg]^{-1} \nonumber ,
\end{eqnarray} 
with 
\begin{equation*} 
\chi_{21} = \frac{\sqrt{3}a^2\omega_{21}^2 d_{21}^2}{4 \pi \hbar  \epsilon_0 N_c \mathcal{V} }.
\end{equation*}
If we want to obtain $c_2(t)$, we have to calculate inverse Laplace transform of $C_2(s)$. This is feasible only numerically.
In figure \ref{f5}  variathions of the real part of $c_2(t)$ is plotted with $d_{21}=100 \ \mathrm{Debye}$, $\lambda_{21} =1.55 \mu \mathrm{m}$, $\Omega_\mathrm{D}= \omega_{21}$, $\delta \kappa = \Gamma \mathrm{K} /10$, $\alpha = 5.38 \times 10^7 \mathrm{m} / \mathrm{s}$, $N_c=7$, $\varphi_\mathrm{a}=0^{\circ}$, $\theta_\mathrm{a}=90^{\circ}$, for two cavity mode volumes ($\mathcal{V}$) and two time intervals. It is seen $c_2(t)$ alternates sinusoidally with a frequency that is inversely proportional to the cavity mode volume and an amplitude which decays very slowly with time. This shows photon is exchanged between the quantum dot and the Dirac modes sinusoidally. By increasing the frequency of this alternation, we hope to enter strong coupling regime and perform some quantum computing algorithms.

\section{Conclusions}\label{s5}

We proposed a new platform to be used for quantum computing. We first showed we can create Dirac cone in the band structure of a PC using coupled cavities inside a triangular lattice. Next we studied the evolution of a quantum dot positioned at the center of one of the cavities due to interaction with Dirac cone modes. We observed the quantum dot exchanged photon with Dirac modes sinusoidally and became entangled with them.

\section*{References}
\bibliography{Ref}

\end{document}